\documentclass[10pt]{article}
\usepackage[T1]{fontenc}
\usepackage[vmargin=2cm]{geometry}
\usepackage{times}
\usepackage{amsmath} 
\usepackage[numbers,sort]{natbib}
\usepackage{multicol}
\usepackage{graphics}
\usepackage{graphicx}
\usepackage{dsfont}
\usepackage{amssymb} 
\DeclareMathOperator*{\fle}{\rightarrow}

\makeatletter
\renewcommand{\fnum@figure}{\small\textbf{\figurename~\thefigure}}
\renewcommand{\fnum@table}{\small\textbf{\tablename~\thetable}}

\newcommand{\be}{\begin{equation}}
\newcommand{\ee}{\end{equation}}

\newcommand{\bi}{\begin{itemize}}
\newcommand{\ei}{\end{itemize}}

\newcommand{\ben}{\begin{enumerate}}
\newcommand{\een}{\end{enumerate}}

\newcommand{\ba}{\begin{array}{lcl}}
\newcommand{\ea}{\end{array}}

\newcommand{\bmat}{\left[\begin{matrix}}
\newcommand{\emat}{\end{matrix}\right]}

\begin{document}

\title{\bf Influence of the number of predecessors in interaction\\ within acceleration-based flow models}
\author{\normalsize Antoine Tordeux\small$^{1,2,}$\footnote{Corresponding author. Email address: \texttt{a.tordeux@fz-juelich.de}}~\normalsize,
Mohcine Chraibi\small$^1$\normalsize, 
Andreas Schadschneider\small$^3$ \normalsize and Armin Seyfried\small$^{1,2}$\\[.25mm]
\footnotesize $^1\,$Institute for Advanced Simulation, Forschungszentrum J\"ulich GmbH, Germany\\[-.75mm]
\footnotesize $^2\,$Computer Simulation for Fire Safety and Pedestrian Traffic, Bergische Universit\"at Wuppertal, Germany\\[-.75mm]
\footnotesize $^3\,$Institut f\"ur Theoretische Physik, Universit\"at zu K\"oln, Germany}%
\date{ }

\maketitle

\begin{abstract}
In this paper, the stability of the uniform solutions is analysed for microscopic flow models in interaction
with $K\ge1$ predecessors.
We calculate general conditions for the linear stability on the ring geometry and 
explore the results with particular pedestrian and car-following models based on relaxation processes. 
The uniform solutions are stable if the relaxation times are sufficiently small. 
However the stability condition strongly depends on the type of models. 
The analysis is focused on the relevance of the number of predecessors $K$ in the dynamics. 
Unexpected non-monotonic relations between $K$ and the stability are presented. 
Classes of models for which increasing the number of predecessors in interaction does not yield an improvement of the stability, or 
for which the stability condition converges as $K$ increases 
(i.e.~implicit finite interaction range) are identified. 
Furthermore, we point out that increasing the interaction range tends to generate characteristic wavelengths in the 
system when unstable.\\[2mm]
{\bf Keywords:} Microscopic flow models, number of neighbors in interaction, homogeneous solution, linear stability analysis
\end{abstract}
\section{Introduction}

\subsection{Context}

Self-driven many-body systems are frequently used to model pedestrian or road traffic streams 
\cite{Chowdhury2000,Helbing2001,Schadschneider2010a}. 
Continuous speed and acceleration models are defined with systems of ordinary, stochastic or delayed differential equations. 
They admit the uniform configurations (where all the spacing and speed are constant and equal) as a stationary solution. 
The stability analysis allows to determine conditions on the parameters 
for which perturbations around the uniform solution disappear (see for instance \cite{Bando1995,Orosz2010,Treiber2013}).
Real observations of pedestrian or vehicle streams present non-homogeneous configurations, 
with propagation of so-called stop-and-go waves for congested density levels \cite{Sugiyama2008,Seyfried2010}.
Therefore, realistic models should have unstable uniform solutions \cite{Orosz2009}.

Traffic flow models can be microscopic, mesoscopic (kinetic), or macroscopic \cite{Treiber2013}. 
Microscopic models describe the individual trajectories in Lagrangian coordinates. 
They are usually called self-driven or motorized particle systems. 
Microscopic models are used since the 1950's for traffic flow modelling in 1D (see for instance \cite{Pipes1953,Chandler1958,Gazis1961}),  
and since the 1970's for pedestrians in 2D (see, e.g., \cite{Hirai1975,Helbing1995,Guo2010,Chraibi2010}). 
Speed-based models refer to as models of the first order while acceleration-based models generally 
refer to as second order models, or again force-based models for pedestrian dynamics in analogy to the Newtonian mechanic. 
The models are investigated theoretically or as simulation tools, 
notably to understand the formation and propagation of stop-and-go waves \cite{Orosz2010,Chraibi2015}. 

The number of neighbors in interaction is a central parameter in microscopic models.  
It is related to as an anticipation factor in traffic modelling (spatial anticipation, see \cite{Treiber2006}). 
Many car-following models with several predecessors in interaction exist in the literature 
(see, e.g., \cite{Bexelius1968,Lenz1999,Nagatani1999,Hoogendoorn2006,Zhu2007,Hu2014}). 
Most of the approaches are extensions of the optimal velocity (OV) model \cite{Bando1995}. 
All of them show that increasing the number of predecessor in interaction stabilizes the uniform solutions.
However, it is unclear whether stabilization occurs for models different from extended OV models. 
Such a feature is fundamental for connected and autonomous vehicles (CAV) \cite{Darbha1999,Wang2004b,monteil2014}. 
For the pedestrian dynamics, the models are generally defined with infinite interaction ranges (i.e.~all neighbors in interaction) 
while finite ones (cut-off radius) are used 
for the simulation to limit the computational effort. 
Nowadays, the impact of the cut-off radius restriction on the dynamics is mainly investigated by simulation (see for instance \cite{KemlohWagoum2012}).
 
Technically, the stability conditions are derived by linearising the models around the homogeneous solution, 
determining the characteristic equation of the linear system and bounding the real part of the characteristic equation solutions. 
In the traffic literature, stability analysis are broadly used with particular models or even variants of existing models 
\cite{Herman1959,Bexelius1968,Bando1995,Nagatani1999,Konoshi2000,Hu2014}. 
To our knowledge, there exists no systematic analysis of generic models, especially when the number of predecessors in interaction is taken as 
an arbitrary parameter. 
Furthermore, the stability analysis are generally investigated for the longest wavelength (see, e.g.,  
\cite{Herman1959,Bexelius1968,Nagatani1999,Hu2014}). 
Such an approach uses a second linearisation process in the characteristic equation allowing to simplify the calculation. 
However the longest wavelength is not systematically the most unstable wave in the system. 
Some studies notably report characteristic wavelengths in stationary states in case of instability (see for instance \cite{Orosz2009}). 

In this article, we calculate general linear stability conditions of an acceleration-based model with $K\ge1$ predecessors in interaction 
for any wavelength. 
The results are explored with particular pedestrian and car-following models including some kind of velocity-relaxation process 
(e.g.~force-based or optimal velocity models). 
The analysis is focused on the relevance of the number of predecessors $K$ for the interaction on the critical relaxation time.  
Unexpected non-monotonic relations between $K$ and the stability are pointed out. 
The organization of the article is the following. 
The model and its uniform solution are defined in the two next subsections. 
The general linear stability conditions are calculated in Sec.~\ref{stab}. 
Applications to distance and speed-based models are investigated in Secs.~\ref{md}, \ref{OVM} and \ref{modgcfm}. 
Secs.~\ref{dis} and \ref{cl} present summary and conclusion. 
 
\subsection{Definition of the model} \label{defmod}

\begin{figure}[!ht]
\begin{center}\vspace{-7mm}
\rotatebox{-90}{\includegraphics[width=.48\textwidth]{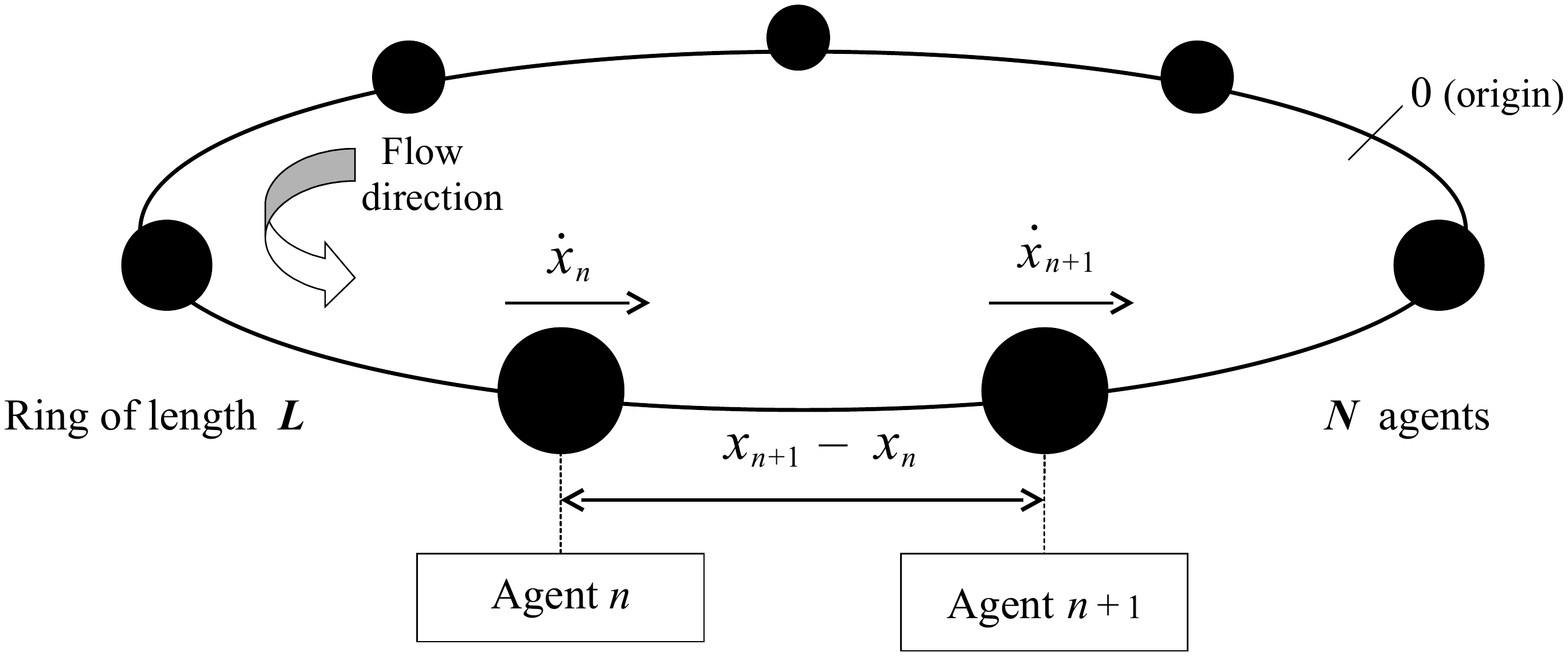}}\vspace{-14mm}
\caption{Scheme of the periodic system with $N$ agents on a ring of length $L$. The predecessor of the agent $n$ is the agent $n+1$ while 
$x_{n+1}-x_n$ is the distance spacing of the agent $n$.}
\label{fig0}\end{center}
\end{figure}

We consider single-file motion with $N$ agents on a one-dimensional ring of length $L$. 
We denote $n$ the index of the agents and $\big(x_1(t),\ldots,x_N(t)\big)$ their curvilinear positions at time $t$ (see Fig.~\ref{fig0}). 
The initial positions are such that 
\be
x_1(0)\le x_2(0)\le\ldots\le x_N(0)\le L+x_1(0).
\ee
The dynamics of the system are supposed strictly asymmetric (i.e. forward interaction). 
They are described by the second order (i.e.~acceleration-based) model
\be
\left\{\ba
\dot x_n(t)&=&v_n(t),\\[1mm]
\dot v_n(t)&=&A\left(v_n(t),x_{n+1}(t)-x_n(t),v_{n+1}(t),\ldots, x_{n+K}(t)-x_n(t),v_{n+K}(t)\right),
\ea\right.
\label{mod}
\ee
for all $t\ge0$ and all $n=1,\ldots, N$. 
The acceleration $A$ of the agent $n$ at time $t\ge0$ depends on the speed and distance spacing, and on the speeds and distance spacings of the 
$K$ predecessors at the same time. 
We assume that $n+k$ is $n+k-N$ if $n+k>N$, and that $x_m-x_n$ is $L+x_m-x_n$ when $m<n$. 
Furthermore we suppose that $0<K<\!\!<N$ and that the function $A$ is at least differentiable. 

\subsection{Uniform solution}\label{soluni}

For a given mean spacing $d=L/N$, we suppose that there exists an equilibrium speed $v$ such that
\be
A\left(v,d,v,2d,v,\ldots, Kd,v,\right)=0.
\label{sol}
\ee 
Under this assumption, the uniform (or homogeneous) configurations $H$ such that
\be
\forall t\ge0,\ \forall n=1,\ldots, N,\qquad x^H_{n+1}(t)-x^H_n(t)=d,\qquad x^H_{n}(t)=x^H_n(0)+vt, 
\label{H}
\ee
are stationary solutions for the system (\ref{mod}). 
Here the uniform configurations $(d,v)$ depend on the initial conditions $\big(x_1(0),\ldots,x_N(0)\big)$. 
We investigate in the next section the linear stability of these solutions. 

\section{Linear stability analysis}\label{stab}

We determine conditions for the parameters of the acceleration-based model~(\ref{mod}) 
for which the uniform solutions (\ref{H}) are linearly stable. 
The local stability analysis, for a finite line of agents with a leader travelling at a known speed, 
is distinguished from the global stability, for agents on a ring or on an infinite line \cite{Orosz2010}. 
The global stability conditions are more restrictive since they contain as well convective perturbations that can locally vanish 
\cite{Treiber2013}. 
Here the global stability conditions are calculated for the ring geometry.

\subsection{Linearisation of the system}

The stability conditions are calculated by considering the evolution of the perturbation 
\be
\tilde x_n(t)=x_n(t)-(x^H_{n}(0)+vt)\qquad\mbox{and}\qquad\tilde v_n(t)=v_n(t)-v.
\ee 
A uniform solution $H$ is said to be stable if $\tilde x_n(t)\fle0$ and $\tilde v_n(t)\fle0$ as $t\fle\infty$ for all $n=1,\ldots, N$. 
By construction $\dot{\tilde x}_n(t)=\tilde v_n(t)$ and, from the Taylor's theorem and by using (\ref{sol}), 
\be\textstyle\dot{\tilde v}_n(t)
= \sum_{k=1}^K\alpha_k\big(\tilde x_{n+k}(t)-\tilde x_n(t)\big)+\sum_{k=0}^K\beta_k\tilde v_{n+k}(t)
+R\big(\tilde v_n(t),\tilde x_{n+1}(t)-\tilde x_n(t),\,\ldots\big),\ee
with 
\be
\alpha_k=\frac{\partial A}{\partial d_k}\left(v_0,d_1,v_1,\ldots,d_k,v_k,\ldots\right),
\qquad\beta_k= \frac{\partial A}{\partial v_k}\left(v_0,d_1,v_1,\ldots,d_k,v_k,\ldots\right),
\label{dp}
\ee
the partial derivatives of $A:(v,d,v,\ldots)\mapsto A(v,d,v,\ldots)$ and where $R(v,d,v,\ldots)$ 
is a function tending to zero as $(v,d,v,\ldots)$ tend to zero. 
Neglecting the rest $R$, we obtain the linear dynamics
\be
\left\{\ba
\dot{\bar x}_n(t)&=& {\bar v}_n(t),\\[1mm]
\dot{\bar v}_n(t)&=&\sum_{k=1}^K\alpha_k\big(\bar x_{n+k}(t)-\bar x_n(t)\big)+\sum_{k=0}^K\beta_k\bar v_{n+k}(t).
\ea\right.
\label{modlin}
\ee

In the following, we focus on the evolution of the variables $(\bar x_n(t),\bar v_n(t))$. 
A uniform configuration $H$ is said to be linearly stable if $\bar x_n(t)\fle0$ and $\bar v_n(t)\fle0$ as $t\fle\infty$ for all $n=1,\ldots, N$.

\subsection{Calculation of the characteristic equation}

The linear stability analysis requires the solution of a second order 
polynomial characteristic equation with complex coefficients (or even a $2N$ polynomial with real coefficients). 
The characteristic equation of the linear system (\ref{modlin}) can be obtained by matrix calculus (see Appendix~1), 
or directly by using the exponential Ansatz 
$\bar x_n(t)=\xi \,e^{\lambda t+in\theta}$ with $\xi\in\mathbb R$ and $\lambda\in\mathbb C$. 
One gets in the last case
\be
\lambda^2=\sum_{k=1}^K\alpha_k(e^{ik\theta}-1)+\lambda\sum_{k=0}^K\beta_ke^{ik\theta}.
\label{eq}
\ee
Here $\theta\in[0,2\pi]$ describes the wavelength of the system when it is unstable. 
We have for the ring $\theta=2\pi l/N$ and $2N$ solutions $(\lambda_l^{(p)}, p=1,2,l=1,\ldots, N)$ for (\ref{eq}).
For $l=N$ (or $\theta=0$), the two solutions are $\lambda^{(1)}_N=0$ and $\lambda^{(2)}_N=\sum_{k=0}^K\beta_k$. 
Assuming 
\be
\sum_{k=0}^K\beta_k<0,
\label{Cp}
\ee
the uniform solution (\ref{H}) is linearly stable if the zeros $(\lambda_l^{(p)}, p=1,2)$ of the characteristic equation (\ref{eq}) 
have strictly negative real parts for all $l=1,\ldots, N-1$. 

\subsection{General stability condition}

Equation~(\ref{eq}) is a second order polynomial of the form $\lambda^2+w_l\lambda+z_l=0$  
with complex coefficients $w_l=\mu_l+i \sigma_l$, $z_l=\nu_l+i\rho_l$ such that
\be
\left|\ba
\mu_l&=&-\sum_{k=0}^K\beta_k\,c_{lk},\\[1mm]
\nu_l&=&\sum_{k=1}^K\alpha_k\left(1-c_{lk}\right),
\ea\right. \qquad\quad
\left|\ba
\sigma_l&=&-\sum_{k=1}^K\beta_k\,s_{lk},\\[1mm]
\rho_l&=&-\sum_{k=1}^K\alpha_k\,s_{lk},
\ea\right.\label{pq}\ee
with $c_{lk}=\cos\left(2\pi lk/N\right)$ and $s_{lk}=\sin\left(2\pi lk/N\right)$. 
The sufficient and necessary conditions for that a polynomial with complex coefficients have all its zeros in the half-plane $\Re(\lambda)<0$ 
are given in \cite[Th.~3.2]{Frank1946}. 
The results are a generalization of the Hurwitz conditions for polynomials with real coefficients. 
These conditions are
\be
\mu_l>0\quad\mbox{and}\quad
\det\bmat
\mu_l&0&-\rho_l\\
1&\nu_l&-\sigma_l\\
0&\rho_l&\mu_l
\emat=\mu_l(\nu_l\mu_l+\rho_l\sigma_l)-\rho_l^2>0,\qquad\forall l=1,\ldots, \lceil N/2\rceil.
\label{CK}
\ee
The condition is general but difficult to interpret. 
It will be investigated with particular distance and speed-based pedestrian and car-following models in the three next sections. 
Note that the inequalities are the same if we substitute $l$ by $N-l$. 
Therefore the study is restricted to $l=1,\ldots, \lceil N/2\rceil$ instead of $l=1,\ldots, N-1$.
The condition (\ref{CK}) is confirmed with some particular models for which the conditions are well-known in the literature, 
see Appendix~2.



\section{Pedestrian models with distance-based repulsive force} \label{md}

Many pedestrian models are described by the sum of a relaxation to the desired speed and additive 
repulsions with the neighbors (see for instance \cite{Helbing1995,Guo2010,Helbing2000, Moussaid2011,Moussaid2012})

\be
\left\{\ba
\dot x_n(t)&=&v_n(t),\\[1mm]
\dot v_n(t)&=&\frac1\tau(v_0-v_n(t))-\sum_{k=1}^K f\left(x_{n+k}(t)-x_n(t)\right),
\ea\right.
\label{mod4}
\ee
with $v_0>0$ the desired speed, $\tau>0$ the relaxation time and $f(\cdot)$ a $C^1$ positive and non-increasing function 
for the repulsion. 
Here, the repulsive force $f(\cdot)$ solely depends on the spacing. 
In this model class, the equilibrium speed $v$ solution of (\ref{sol}) is, for a given mean spacing $d$,
\be
v=v_0-\tau\sum_{k=1}^Kf(kd).
\ee
The equilibrium speed 
is positive if $\tau\le v_0/\sum_kf(kd)$ and converges as $K\rightarrow\infty$ if $\sum_kf(kd)<\infty$. 
This is a preliminary restriction on the relaxation time parameter that a model has to satisfy. 
Note that long-range models cannot by construction fulfil the positivity condition of the equilibrium speed as $K\fle\infty$. 

With the model~(\ref{mod4}), the partial derivatives are $\beta_0=-1/\tau$, $\beta_k=0$ for all $k>0$, and $\alpha_k=-f'(kd)$ 
for all $k$.
The parameters (\ref{pq}) are 
\be\left|\ba
\mu_l&=&1/\tau,\\[1mm]
\nu_l&=&-\sum_{k=1}^Kf'(kd)\left(1-c_{lk}\right),
\ea\right.\qquad\left|\ba
\sigma_l&=&0,\\[1mm]
\rho_l&=&\sum_{k=1}^Kf'(kd)s_{lk},
\ea\right.\ee
and do not depend on the equilibrium speed $v$.
The first  linear stability condition in (\ref{CK}) is $-1/\tau<0$. 
It is always true and implies the preliminary assumption (\ref{Cp}). 
The second condition in (\ref{CK}) gives the stability threshold for the relaxation time
\be
0<\tau<\tau_K=\min_{l=1,\ldots, \lceil N/2\rceil}\tau_K^{(l)},\qquad\mbox{with}\quad \tau_K^{(l)}=
\frac{\sqrt{-\sum_{k=1}^Kf'(kd)\left(1-c_{lk}\right)}}{\left|\sum_{k=1}^Kf'(kd)s_{lk}\right|}.
\label{critt}
\ee
Note that $f'(d)\le0$ for all $d$. 
The stability occurs for a relaxation time $\tau$ sufficiently small. 
Since $f$ is $C^1$ positive and non-increasing, $\lim_{x} f(x)=f(y)- \int_y^\infty |f'(u)|\,\mbox{d}u=0$ for all $y>0$. 
This implies $\int_y^\infty |f'(u)|\;\mbox{d}u<\infty$ or again, using the Cauchy criterion and changing the variable, 
\be \sum_{k=1}^\infty |f'(dk)|<\infty\qquad\mbox{for any } d>0.\ee
This proves the convergence of $\tau_K^{(l)}$ and $\tau_K$ within pedestrian models with distance-based repulsive force.
Therefore the impact on the stability of the number of predecessors in interaction $K$ is limited, 
i.e.~there exists an intrinsic interaction range, as soon as the repulsion $f(x)\fle 0$ as $x\fle\infty$ 
(irrespective of the convergence speed).   
The stability condition at the limit $K\rightarrow\infty$ may be well approximated for a finite value of $K$. 
Therefore the use of cut-off radius in such models can have not effect on the dynamics. %
Yet, the value of the range depends on the convergence speed of the series $\sum_k |f'(dk)|$. 
This point is investigated using exponential and algebraic repulsive forces. 
\paragraph{Exponential force}
We first consider the exponential repulsive force with parameters $A,B>0$ 
\be
f(d)=Ae^{-d/B}.
\label{exp}
\ee
This short-range force is used within the social force model \cite{Helbing1995}. 
The equilibrium speed is positive with this model as soon as 
\be
\tau\le \frac{v_0}{A\sum_{k=1}^Ke^{-kd/B}}.
\ee
In the following, we use the dimensionless spacing and critical relaxation time
\be
u=d/B\qquad\mbox{and}\qquad\tilde\tau^{(l)}_K=\sqrt{\frac AB}\tau_K^{(l)}.
\label{tT}
\ee
We have with the exponential repulsive force (\ref{exp})
\be
\tilde\tau_K^{(l)}(u)=
\frac{\sqrt{\sum_{k=1}^Ke^{-ku}\left(1-c_{lk}\right)}}{\left|\sum_{k=1}^Ke^{-ku}s_{lk}\right|}.
\label{ct1}
\ee
Here $u$ is a form parameter and $A$ and $B$ are scale parameters for the dimensioned $\tau_K^{(l)}$, while $v_0$ has no influence. 

\paragraph{Algebraically decaying force}
We consider the algebraic repulsive force with parameters $A,B,q>0$ 
\be
f(d)=\frac A{(d/B)^q}.
\label{pol}
\ee
This model is used in \cite{Guo2010} with $q=1$ and in \cite{Helbing2000} with $q=2$. 
Here, the repulsive force can be short or long range according to the value of $q$. 
The model is long range for $q\le1$ in one dimension. 
The equilibrium speed converge as $K\rightarrow\infty$ if and only if $q>1$ (i.e.~for short range forces). 
More precisely, the equilibrium speed is positive if 
\be
\tau\le \frac{u^qv_0}{A\sum_{k=1}^Kk^{-q}}.
\ee
The dimensionless critical relaxation time is
\be
\tilde\tau_K^{(l)}(u,q)=\frac{\sqrt{\frac{u^{q+1}}q
\sum_{k=1}^Kk^{-q-1}(1-c_{lk})}}{\left|\sum_{k=1}^K k^{-q-1}s_{lk}\right|}.
\label{ct2}
\ee 
Here $A$ and $B$ are scale parameter for $\tau_K^{(l)}$, $v_0$ has no influence while $u$ and $q$ are form parameters.

\subsection{Stability condition function of $K$}

The uniform solutions (\ref{H}) are linearly stable for the distance-based models with forces (\ref{exp}) and (\ref{pol}) 
if the relaxation time $\tau$ is strictly less than critical time $\tau_K=\min_l\tau_K^{(l)}$ 
(i.e.~by using dimensionless variables, if  $\sqrt{A/B}\tau<\tilde\tau_K$). 
Thus we have to calculate the minimum of the functions $l\mapsto\tilde\tau_K{(l)}$ to determine the stability condition. 
Yet, the sign of the derivative of the function $l\mapsto\tilde\tau_K{(l)}$ is hard to extract. 
Therefore, we investigate it numerically with $N=10^5$. 

\begin{figure}[!ht]
\begin{center}
\includegraphics[width=.95\textwidth]{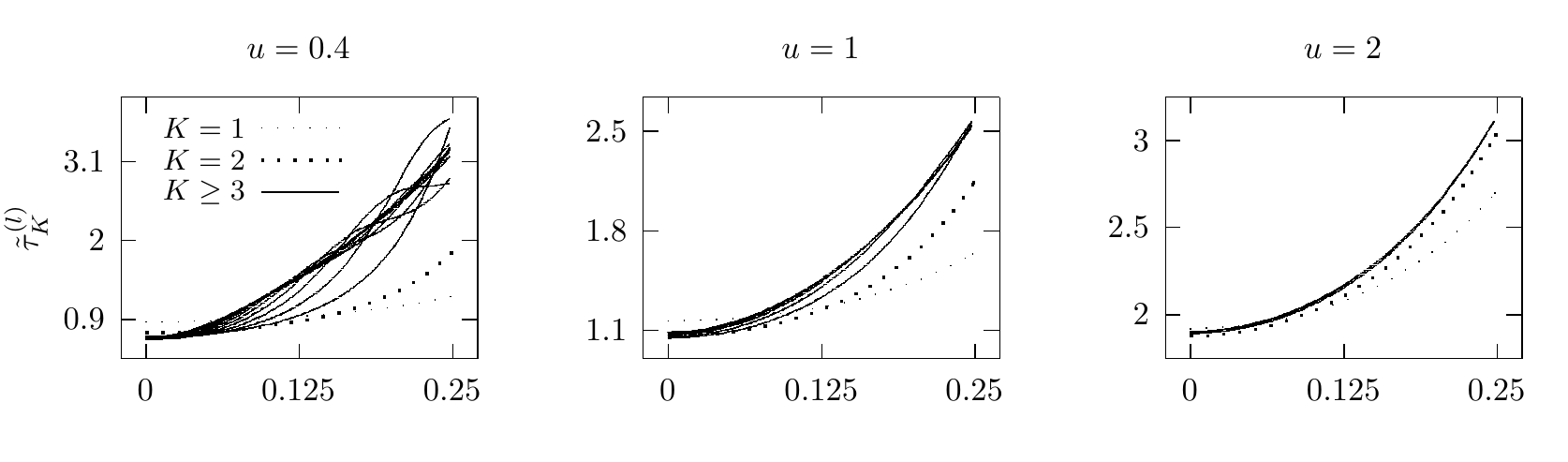}\\[-3mm]
\includegraphics[width=.95\textwidth]{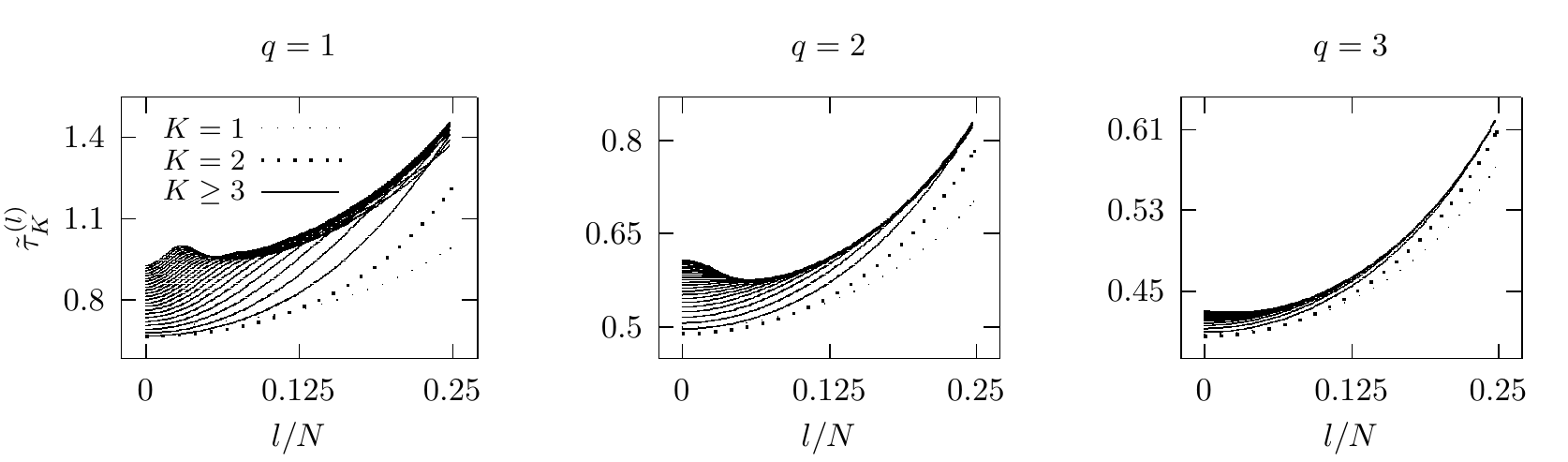}\vspace{-2mm}
\caption{Dimensionless critical relaxation time as a function of the wavelength $l$ for $K=1,\ldots, 25$. 
Top panels: Exponential force~(\ref{exp}), $u=0.4,1,2$. 
Bottom panels: Algebraic force~(\ref{pol}), $q=1,2,3$ and $u=1$. 
The stability holds when the relaxation time is smaller than the critical time. 
For the exponential model the most unstable wavelength is the longest one for which $l=1$.  
For the algebraic model the most unstable wavelength has a characteristic size when the number of predecessors in interaction $K$ 
is sufficiently high.}
\label{stabl}
\end{center}
\end{figure}

The critical times (\ref{ct1}) for the exponential model are plotted  as a function of $l$ in Fig.~\ref{stabl}, top panels. 
Here, $K$ varies from $1$ to $25$, and we compare $u=0.4$, $1$ and $2$. 
We observe that for the exponential force, the $\tilde\tau_K^{(l)}$ are minimal for $l=1$ for all $K$, i.e. 
$\tilde\tau_K=\tilde\tau_K^{(1)}$. 
Further numerical investigations (not shown here) confirm it. 
This means that the longest wavelength is always the most unstable with the exponential repulsive force.
The critical times (\ref{ct2}) for the algebraic model are plotted  as a function of $l$ in Fig.~\ref{stabl}, bottom panels, 
with $q=1$, $2$ and $3$. 
One observes that $\tau_K^{(l)}$ are minimal for $l=1$ when $K$ is low. 
For high $K$, the minimums are reached for $l_{q,K}\propto N$. 
Further results show that $l_{q,K}$ converges when $K$ increases. 
Therefore the most unstable wavelength has a characteristic size proportional to $N$ with the algebraic force, 
if $K$ is sufficiently high. 

\begin{figure}[!ht]
\begin{center}
\includegraphics[width=.95\textwidth]{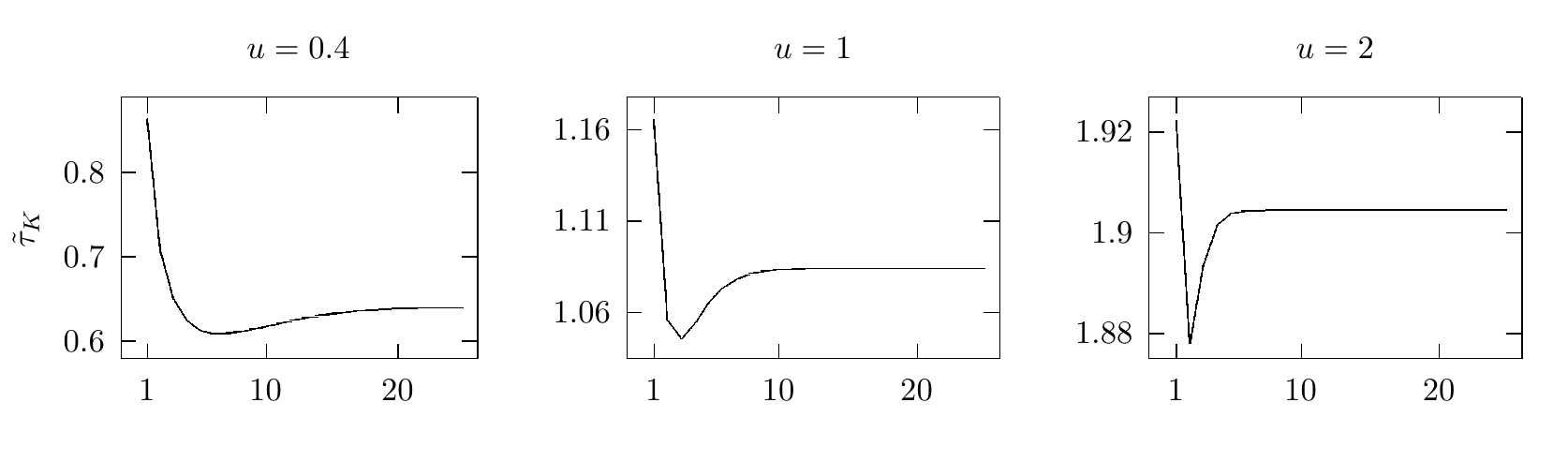}\\[-3mm]
\includegraphics[width=.95\textwidth]{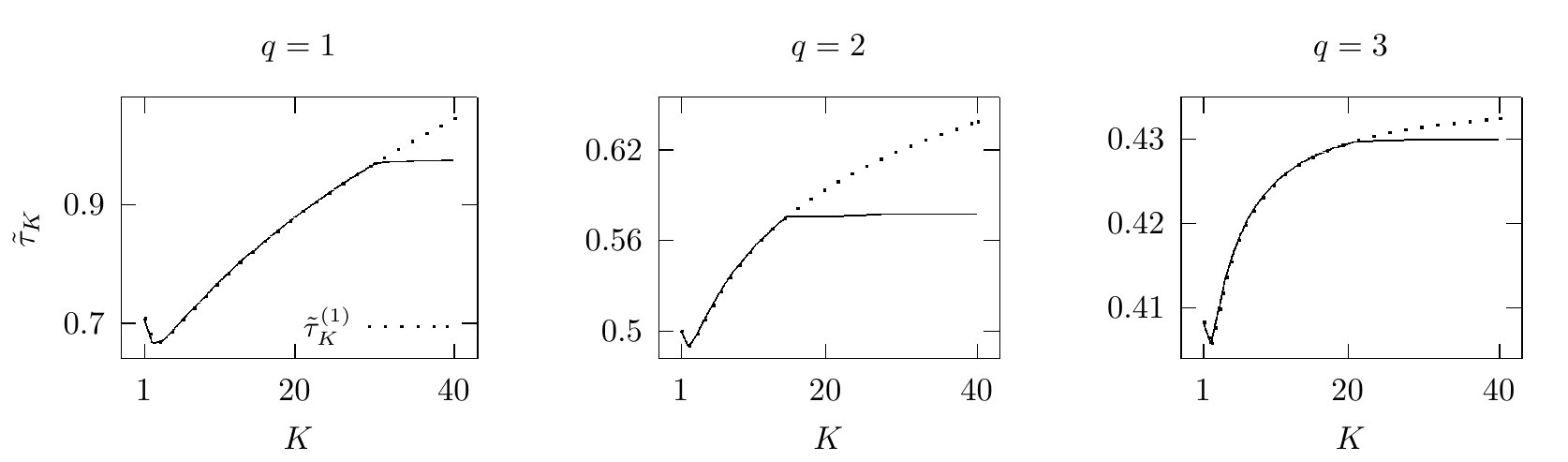}\vspace{-2mm}
\caption{Dimensionless critical relaxation time $\tilde\tau_K=\min_l\tilde\tau_K^{(l)}$ 
as a function of $K$ for the exponential force, $u=0.4,\ 1,\ 2$, top panels, and for 
the algebraic force, $q=1,\ 2,\ 3$ and $u=1$, bottom panels.
The stability holds when the relaxation time is smaller than the critical time. 
Surprisingly a non-monotonic relation is observed between the number of predecessors in interaction and the stability for both 
exponential and algebraic models.}
\label{stabK}
\end{center}
\end{figure}

The critical relaxation times $\tilde\tau_K=\min_l\tilde\tau_K^{(l)}$ determine the border of the linear stability 
for any wavelength. 
They are plotted as function of $K$ in Fig.~\ref{stabK} for the exponential force (top panels), 
and for the algebraic one (bottom panels). 
The critical time converges to a constant value through a single damped oscillation for both models. 
This non-monotonic relation between $K$ and the stability is unexpected. 
Increasing the number of pedestrians in interaction first results in a decrease of the stability (at list until $K=2$). 
Then increasing $K$ increases $\tau_K$ and so the stability. 
The convergence of $\tau_K$ is relatively smooth for the exponential force. 
One observes a brusque transition for the algebraic force when the stability is broken for wavelength smaller than the longest one.  
In general, the speed of damping of the function $K\mapsto\tilde\tau_K$ depends on the spacing parameter $u$ 
for the exponential model, and on the force range parameter $q$ for the algebraic model.

\subsection{Variation of the stability condition}

\begin{figure}[!ht]
\begin{center}
\includegraphics[width=.633\textwidth]{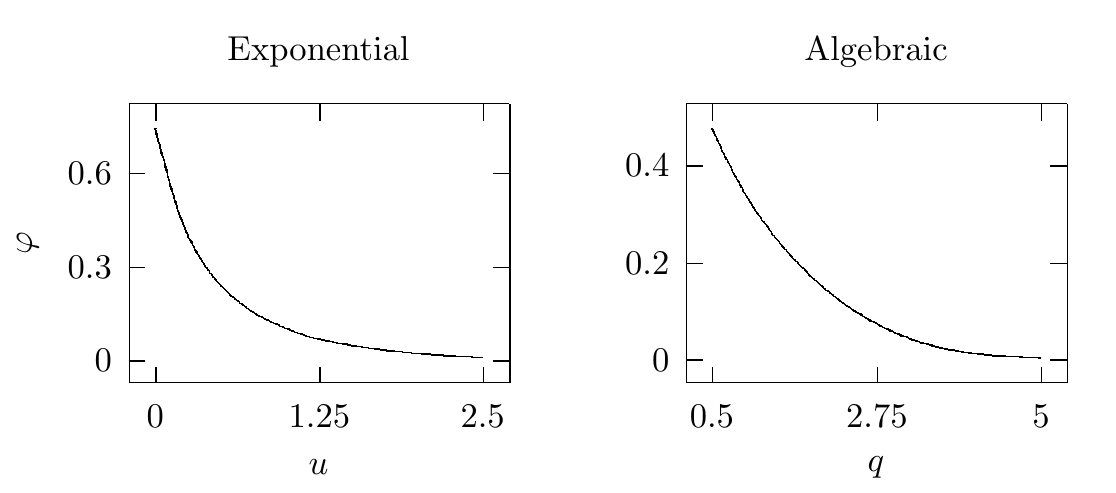}\vspace{-2mm}
\caption{Proportion of variation $\varphi$ of the critical relaxation time as a function of the dimensionless spacing $u$  
(exponential force, left panel), and as a function of the range parameter $q$ (algebraic model, right panel). 
For the exponential model the stability varies for high density (i.e.~low $u$). 
For the algebraic model it varies for long range forces (i.e.~low $q$) independently of the density.}\vspace{-5mm}
\label{figuq}
\end{center}
\end{figure}

The proportion of variation of the critical relaxation time
\be
\varphi=1-\frac{\min_K \tilde\tau_K}{\max_K \tilde\tau_K}=
1-\frac{\min_K \tau_K}{\max_K \tau_K},
\label{prop}
\ee
allows to quantify the influence of the number of predecessors in interaction $K$ on the stability. 
It is the same for initial and dimensionless critical times $\tau_K$ and $\tilde\tau_K$.
For $\varphi\approx0$, the model weakly depends on the interaction range and inversely for $\varphi\approx1$. 
The proportion $\varphi$ does not depend on $A$ and $B$ for both exponential and algebraic forces. 
It depends on $u$ for the exponential model, while it depends on $q$ for the algebraic model, but not on $u$. 
In the Fig.~\ref{figuq}, the variation $\varphi$ is plotted as a function of $u$ for the exponential model (left panel), 
and as a function of $q$ for the algebraic model (right panel).
 $\varphi$ tends to zero as the dimensionless mean spacing $u$ increases with the first model.  
This means that $\tilde\tau_K$ marginally varies for low density levels. 
The same phenomenon occurs as $q$ increases with the algebraic model. 

\subsection{Stability conditions for other parameters}

\begin{figure}[!ht]
\begin{center}
\includegraphics[width=.633\textwidth]{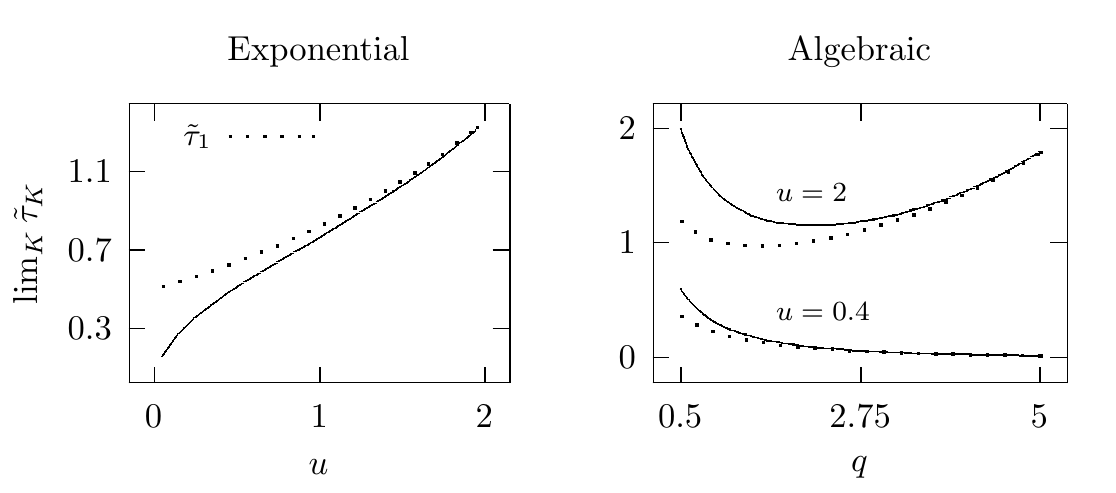}\vspace{-2mm}
\caption{Dimensionless critical relaxation time for $K=1$ (dotted lines), 
and at the limit $K\rightarrow\infty$ (continuous lines). 
Left: Critical time as a function of $u$ (exponential force). 
Right: Critical time as a function of $q$ for $u=0.4$ and $u=2$ (algebraic force). 
As expected (see Fig.~\ref{figuq}), the critical times converge to the same value as, respectively, $u\fle\infty$ and $q\fle\infty$. 
The stability is improved as the spacing increases for the exponential model, while 
the stability is negatively impacted by the force range $q$ if $u<1$ and admits a minimum if $u>1$ for the algebraic model.}
\label{figuqb}
\end{center}
\end{figure}

The critical relaxation time $\tilde\tau_K$ depends on the parameter $u$ for the exponential force, see (\ref{ct1}), and on $(u,q)$ for 
the algebraic force, see (\ref{ct2}). 
We investigate here how the stability condition depends on these parameters. 
For the exponential model, the function $u\mapsto\tilde\tau_K(u)$ increases as $u$ increases. 
This means that the stability improves as the distance spacing increases, for any $K$.
One has
\be
\frac{\partial \tilde\tau_K^{(l)}}{\partial u}(u)
=\frac{\tilde\tau_K^{(l)}(u)}2g_K(u)>0
\ee
with $g_k(u)\ge1$ for all $u$, $l$ and $K$. 
More precisely we have $g_1(u)=1$ for all $u$ and $l$, while $\lim_ug_K(u)=1$ for all $l$ and $K>1$. 
In Fig.~\ref{figuqb}, left panel, the increasing critical time $\tilde\tau_K(u)$ at the limit $K\rightarrow\infty$ 
is compared to the time $\tilde\tau_1(u)$ for $K=1$. 
One has $\lim_K\tilde\tau_K(u)<\tilde\tau_1(u)$ for all $u$, while, as expected since the variation tends to zero,  
$\lim_u\tilde\tau_1(u)=\lim_u\tilde\tau_K(u)$ for any $K$. 
For the algebraic force, the function $u\mapsto\tilde\tau_K(u,q)$ increases as $u$ increases since
\be
\frac{\partial \tilde\tau_K^{(l)}}{\partial u}(u,q)=\frac{q+1}u\tilde\tau_K^{(l)}(u,q)>0.
\ee 
The relation $q\mapsto\tilde\tau_K^{(l)}(u,q)$ is more complicated. 
One has
\be
\frac{\partial \tilde\tau_K^{(l)}}{\partial q}(u,q)=\frac{\tilde\tau_K^{(l)}(u,q)}2\left(\ln u-\frac1q+h_K(q)\right).\\[2.5mm]
\ee
For $K=1$,  $h_1(q)=0$ for all $q$, and the sign of $\partial \tilde\tau_K^{(l)}/\partial q$ is the sign of $\ln u-1/q$. 
It is negative for all $q$ if $u<1$ while it is minimum for $q=1/\ln u$ if $u>1$. 
Comparable properties are obtained for $K>1$ (see Fig.~\ref{figuqb}, right panel). 

\section{Optimal velocity model}\label{OVM}

The multi-anticipative optimal velocity model \cite{Lenz1999} with $K\ge1$ predecessors in interaction is
\be
\left\{\ba
\dot x_n(t)&=&v_n(t),\\[1mm]
\dot v_n(t)&=&\sum_{k=1}^K a_k\left\{V\left(\frac1k (x_{n+k}(t)-x_n(t))\right)- v_n(t)\right\}.
\ea\right.
\label{mod3}
\ee
The equilibrium speed is given by the function $V$ for this model. 
We assume in the following that $V'(d)>0$ and $1/a_k=\tau k^q$ with $\tau>0$ the relaxation time with the first predecessor and 
$q\ge0$ a parameter calibrating the range of the force (in a similar way than the algebraic force).
The stability condition is
\be
0<\tau\cdot V'(d)<\frac{\Big(\sum_{k=1}^Kk^{-q}\Big)^2
\sum_{k=1}^K\left(1-c_{lk}\right)k^{-(q+1)}}{\left(\sum_{k=1}^Ks_{lk}k^{-(q+1)}\right)^2}=:\tilde\tau_K^{(l)}.
\label{CKovmb}
\ee
The mean spacing has only a role through the derivative of the optimal speed function that is a scale parameter. 
Only the parameter $q$ is a form one. 

\subsection{Stability condition function of $K$}
\begin{figure}[!ht]
\begin{center}\vspace{5mm}
\includegraphics[width=.633\textwidth]{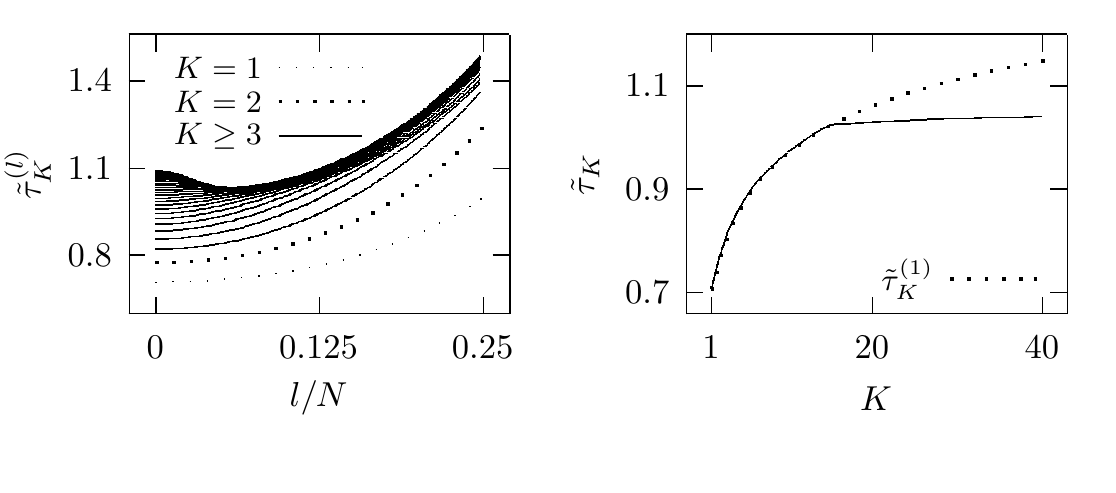}\vspace{-7mm}
\caption{Dimensionless critical relaxation time for the OV model with $q=2$. 
The stability holds when the relaxation time is smaller than the critical time. 
Left: Critical time as a function of the wavelength $l$ for $K=1,\ldots,25$.  
As for the algebraic, the most unstable wave has a characteristic size when $K$ is sufficiently high. 
Right: Critical time as a function of the number of predecessors in interaction $K$.  
By opposition to exponential and algebraic models, the stability is always improved when $K$ increases. 
Such a feature is classically observed in the literature \cite{Bexelius1968,Lenz1999,Nagatani1999,Hoogendoorn2006,Zhu2007,Hu2014}.}
\label{figovm}
\end{center}
\end{figure}

The critical times are proportional to the square of $\sum_kk^{-q}$ and converges if and only if $q>1$. 
Oppositely to the force-based model~(\ref{mod4}), the critical times not always converge as $K$ increases within the OV model. 
When $q>1$, the forms of the functions $l\mapsto\tilde\tau_K^{(l)}$ are comparable to the ones obtained with the algebraic force 
(see Fig.~\ref{figovm}, left panel, and Fig.~\ref{stabl}, bottom middle panel). 
The critical time does not describe damped oscillation with the OV model (see Fig.~\ref{figovm}, right panel). 
This means that the stability improves as the number of predecessors in interaction $K$ increases for any $V'(d)>0$ and $q>0$. 
This in agreement with the results presented in the literature about extended OV models 
\cite{Bexelius1968,Lenz1999,Nagatani1999,Hoogendoorn2006,Zhu2007,Hu2014}.
The critical times converge if $q>1$. 
In this case, as for the previous pedestrian models, the impact of $K$ on the stability is limited (see also \cite{Hasebe2004}).
Note that as previously the critical relaxation times are not minimal for the longest wavelength $l=1$ if $K$ is sufficiently large 
(e.g.~for approximately $K>17$ in Fig.~\ref{figovm}, right panel). 

\subsection{Variation of the stability condition}
\begin{figure}[!ht]
\begin{center}\vspace{5mm}
\includegraphics[width=.633\textwidth]{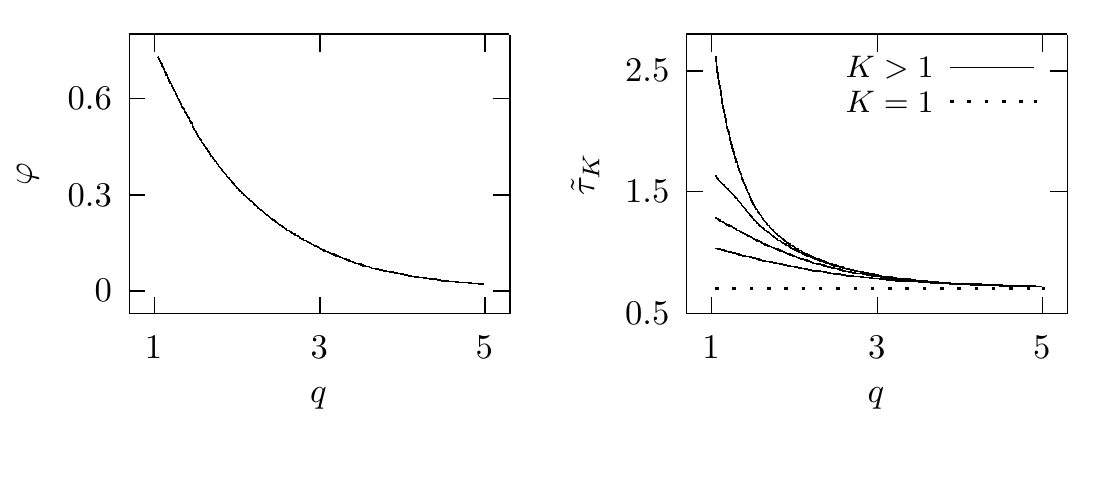}\vspace{-7mm}
\caption{Left: Proportion of variation $\varphi$ of the critical relaxation time for the OV model. 
As for the algebraic model, see Fig.~\ref{figuq}, right panel, the influence of the number of predecessors in interaction $K$ 
decreases as the model becomes short range (i.e.~as $q$ increases). 
Right: Dimensionless critical time for $K=1,5,10,20$ and at the limit  $K\rightarrow\infty$. 
The stability is always improved as $K$ increases, yet the effect is limited by the range $q$.}
\label{figovmp}
\end{center}
\end{figure}

The proportion of variation $\varphi$ of the critical time (see Eq. (\ref{prop})), 
is not defined when $q\le1$ with the OV model~(\ref{mod3}) since in this case $\tilde\tau_K$ diverges. 
For $q>1$, the proportion tends to zero as $q$ increases (see Fig.~\ref{figovmp}, left panel). 
As expected, and as we observe with the algebraic model (see Fig.~\ref{figuq}, right panel) the influence of $K$ 
decreases as the model becomes short range (i.e.~as $q$ increases). 
The critical time does not depends on $q$ for $K=1$. For any $K>1$, the times decreases as $q$ increases. 
Therefore the stability is negatively influenced by the range $q$ (see also \cite{Hasebe2003}). 
The constant minimal value for $K=1$ corresponds here to the limit 
as $q$ increases for any $K$ (see Fig.~\ref{figovmp}, right panel). 
This means that, oppositely to the algebraic model with $u<1$, the OV model can remain stable at 
the limit $q\rightarrow\infty$.

\section{Generalized centrifugal force model}\label{modgcfm}

The repulsive force in the generalized centrifugal force (GCF) model \cite{Chraibi2010} depends, as in the extended 
social force model given in \cite{Gao2013}, on the distance spacings and also on the speeds of the predecessors.
The GCF model is 
\be
\left\{\ba
\dot x_n(t)&=&v_n(t),\\[-1mm]
\dot v_n(t)&=&\frac1\tau(v_0-v_n(t))-
\sum_{k=1}^K\frac{\displaystyle\big(\eta v_0+\delta(v_n(t)-v_{n+k}(t))\big)^2}{\displaystyle x_{n+k}(t)-x_n(t)-h(v_n(t))-h(v_{n+k}(t))}.
\ea\right.
\label{mod5}
\ee
with $v_0,\tau,\eta>0$ and $h:v\mapsto a+b v$ with $a,b\ge0$ parameters for the size of the pedestrian. 
Here $\delta$ is a boolean variable for the speed difference term. 
The model is the centrifugal force one \cite{Yu2005} if $a=b=0$ and $\delta=1$, 
while it is the previous algebraic model~(\ref{pol}) with $q=1$ if $\delta=a=b=0$.


The equilibrium speed $v$ solution of (\ref{sol}) is not explicit with the GCF model.
For a given mean spacing $d$, it is the solution of
\be
g(v)=:\frac1\tau(v_0-v)-\sum_{k=1}^K\frac{\left(\eta v_0\right)^2}{kd-2h(v)}=0.
\ee
Here  
$g(v) \fle\infty$ as $v\rightarrow -\infty$ and $g(v)=-\infty$ as $v\rightarrow \infty$.  
Moreover $g(v)=-\infty$ as $v\rightarrow v_k^-$ and 
$g(v)=\infty$ as $v\rightarrow v_k^+$, for $v_k(d)=(kd/2-a)/b$ and $k=1,\ldots, K$. 
The function $g$ is continuous and strictly decreasing on the subsets $(-\infty,v_1)$, $(v_{1},v_2)$,\ldots, $(v_{K-1},v_K)$, $(v_K,\infty)$. 
From the opposite signs of the limits, we deduce by continuity that $g(\cdot)$ admits $K+1$ solutions, one on each subset.
We want that the solution $v$ satisfies $d-2h(v)>0$. 
It is only the case for the solution $v$ belonging to the subset $(-\infty,v_1(d))$. 
We solely consider in the following this solution. 
It depends on the $v_0,K,d,\tau$, $a$ and $b$ parameters. 
The solution is one of the roots of a $K+1$ polynomial equation. 
It has no explicit definition as soon as $K>1$.

\subsection{Stability condition function of $K$}

We calculate the stability condition (\ref{CK}) within the GCF model.
Denoting 
\be
A_k(v)=\eta v_0/(kd-2h(v))>0,
\ee 
we have $\beta_0=-\frac1\tau-\sum_{k=1}^KA_k(2+bA_k)$, $\beta_k=A_k(2-bA_k)$, and $\alpha_k=A_k^2$ for all $k=1,\ldots, K$. 
The preliminary condition (\ref{Cp}) as well as the first condition in (\ref{CK}) hold. 
The parameters (\ref{pq}) are 
\be\left|\ba
\mu_l=\frac1\tau+2\sum_{k=1}^KA_k(1-c_{lk})+\tau_r\sum_{k=1}^KA_k^2(1+c_{lk}),\\[1mm]
\nu_l=\sum_{k=1}^KA_k^2(1-c_{lk}),\ea\right.\quad\left|\ba 
\sigma_l=-\sum_{k=1}^KA_k(2-bA_k)s_{lk},\\[1mm]
\rho_l=-\sum_{k=1}^KA_k^2s_{lk}.\ea\right.\ee 
Here the parameters depend on the equilibrium speed $v$ and therefore on $\tau$. 
Since the analytical form of $v$ is unknown, we can not explicitly express the stability condition on $\tau$ 
as we did previously. 
The critical relaxation times $\tau_K$ are investigated numerically with $N=10^5$. 
The results show that the linear stability occurs if the parameter $\tau$ is smaller than $\tau_K$ and 
inversely. 
As before, the stability requires a sufficiently fast relaxation to the desired speed.

\begin{figure}[!ht]
\begin{center}
\includegraphics[width=.95\textwidth]{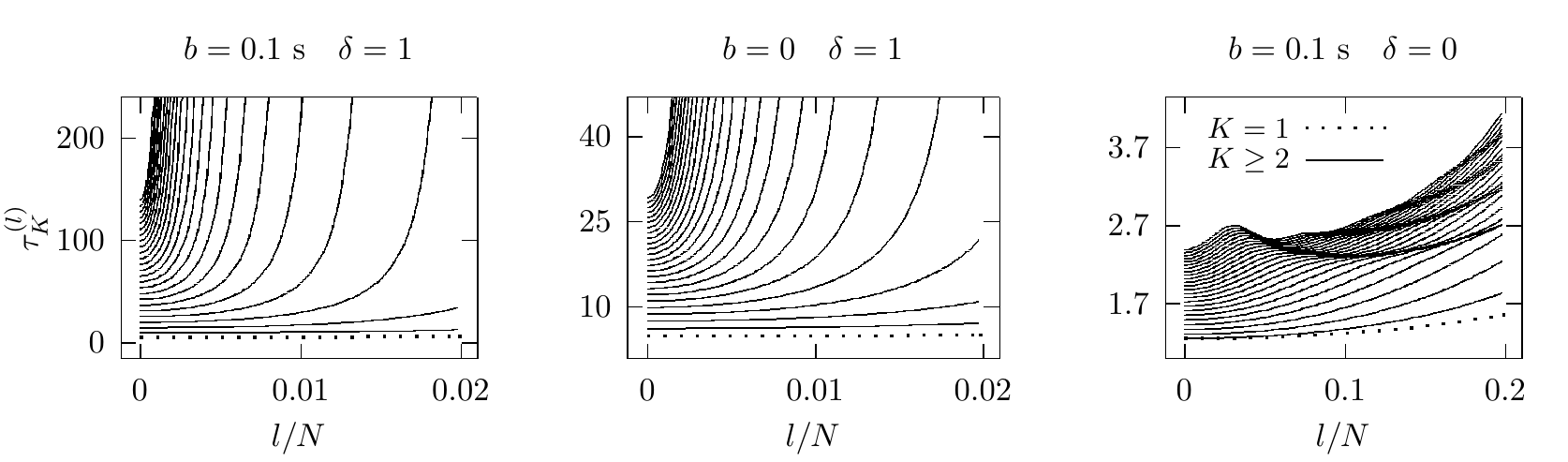}\\[-3mm]
\includegraphics[width=.95\textwidth]{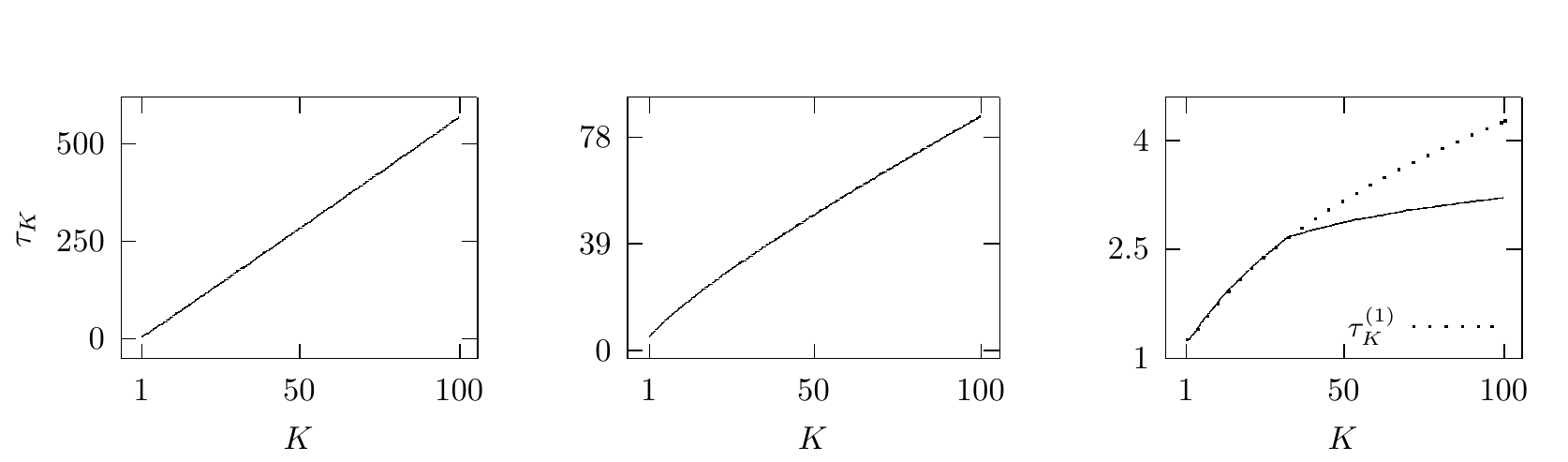}
\caption{Critical relaxation time (in seconds) for the GCF model with $d=1$~m.  
The stability holds when the relaxation time is smaller than the critical time. 
Top panels: Critical time as a function of the wavelength $l$ for $K=1,\ldots, 25$. 
The most unstable wave has a characteristic size only if the speed difference is not taken into account (i.e.~for $\delta=0$).  
Bottom panels: Critical time $\tau_K=\min_l\tau_K^{(l)}$ as a function of $K$. 
The critical time does not converge when the speed difference term is considered (i.e.~for $\delta=1$) while it converges for 
$\delta=0$.}
\label{stabgcfm}
\end{center}
\end{figure}

The critical times as a function of $l$ are plotted Fig.~\ref{stabgcfm}, top panels, for $K=1,\ldots,25$. 
Here, the values of the parameters are: $\eta=0.3$ and $v_0=1.5$~m/s (see \cite{Chraibi2010}). 
We use $a=0$ in order to obtain the algebraic distance-based model at the limit $b\rightarrow0$ if $\delta=0$. 
We compare the cases $b=0.1$~s and $\delta=1$, $b=0$ and $\delta=1$, and $b=0.1$~s and $\delta=0$. 
The results show that the critical time is minimum for $l=1$, for all $K$ if $\delta=1$, while the critical time is 
minimum for $l>\!>1$ if $K$ is large when $\delta=0$. 
As for the exponential model, the longest wavelength is the most unstable if the speed difference term is considered. 
If $\delta=0$, as for the algebraic and OV models, specific wavelengths are the most unstable if $K$ is sufficiently high. 

The critical time for any wavelength $K\mapsto\tau_K$ are drawn in Fig.~\ref{stabgcfm}, bottom panels. 
When $\delta=1$, $\tau_K$ linearly increases with $K$, and does not converge (further experiments done with higher $K$ confirm it). 
The stability always occurs if enough predecessors are taken into account. 
Furthermore, the speed at which $\tau_K$ diverges is positively correlated to $b$, i.e.~increasing $b$ increases the stability. 
When the speed difference term is deleted, i.e.~for $\delta=0$, $\tau_K$ converges (see Fig.~\ref{stabgcfm}, bottom right panel, 
here again further experiments done with higher $K$ confirm it). 
In this last case, the impact of the parameter $b$ on the stability is not straightforward. 
It can be positive or negative according to the value of $K$ or mean spacing $d$.

\subsection{Stability condition for other parameters}

\begin{figure}[!ht]
\begin{center}
\includegraphics[width=.633\textwidth]{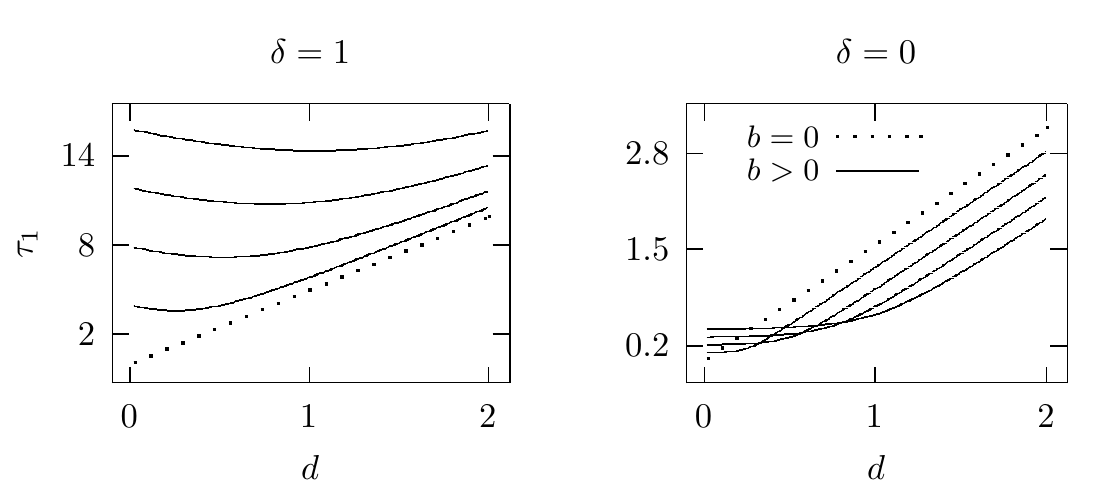}\\[-3mm]
\includegraphics[width=.95\textwidth]{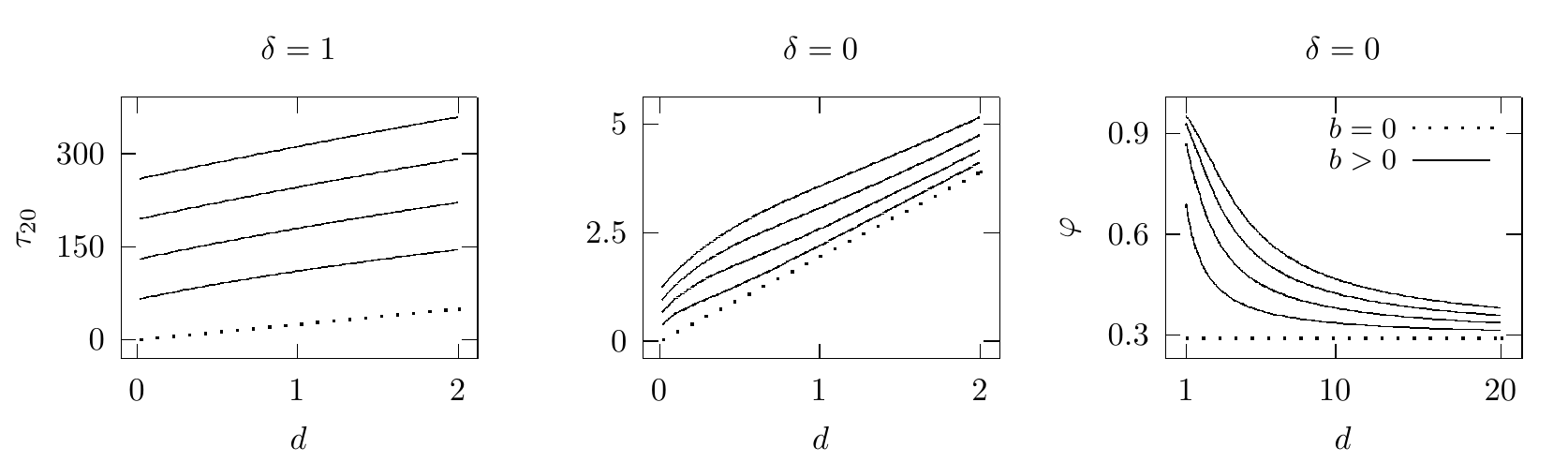}\vspace{-2mm}
\caption{Critical relaxation time (in seconds) as a function of the mean distance spacing $d$ (in meters) 
for the GCF model with size slope $b$ varying from $0$ to $0.4$~s by step of $0.1$~s. 
Top panels: $K=1$ predecessor in interaction. 
The parameter $b$ positively impacts the stability when the speed difference term is taken into account ($\delta=1$, top left panel), 
while it negatively impacts the stability if $\delta=0$ when the spacing is sufficiently high (top right panel). 
Bottom left and middle panels: $K=20$ predecessors in interaction. 
Here, the size slope $b$ always positively impacts the stability. 
Bottom right panel: Proportion of variation of the stability condition (see Eq.~(\ref{prop})). 
The parameter $b$ mainly influences the stability condition when the spacing is low (i.e.~for congested states). }
\label{figgcfmmd}
\end{center}
\end{figure}

The stability condition (\ref{CK}) is investigated with the GCF model as a function of the mean spacing $d$ and size slope 
$b$ parameters. 
As for the distance-based models (\ref{exp}) and (\ref{pol}), the stability always holds as the density tends to zero 
(see Fig.~\ref{figgcfmmd}). 
Yet, the role of $b$ differs according to $d$, $K$ and $\delta$.
When $\delta=1$, the parameter $b$ positively impacts the stability, i.e.~$\tau_K$ increases as $b$ increases, for any $K$. 
For $K=1$ and $b$ sufficiently high, the function $d\mapsto\tau_k(d)$ admits a minimum (see Fig.~\ref{figgcfmmd}, top left panel). 
This feature disappears when $K$ increases (see Fig.~\ref{figgcfmmd}, bottom left panel), the critical time tending 
to be concave. 
Different characteristics are obtained when the speed difference term is not taken into account, i.e.~for $\delta=0$. 
When $K$ is small, the parameter $b$ positively influences $\tau_K$ for small spacing $d$ (see Fig.~\ref{figgcfmmd}, top right panel). 
When $d$ is high (approximately for $d>1$~m in the example), the stability decreases as $b$ increases. 
When the number $K$ of predecessors in interaction is high (approximately higher that $8$ here), 
the parameter $b$, as for the case $\delta=1$, always positively impacts the stability (see Fig.~\ref{figgcfmmd}, bottom middle panel).

The proportion of variation of the critical time $\varphi$ (see Eq.~(\ref{prop})  
has only a sense if the speed difference term is not taken into account, i.e.~for $\delta=0$ (since $\tau_K$ diverges for $\delta=1$). 
Only the parameters $d$ and $b$ are form ones for $\varphi$. 
It decreases as the spacing $d$ increases (see Fig.~\ref{figgcfmmd}, bottom right panel). 
This means that the influence of $K$ on the stability decreases as the density decreases. 
When $b=0$, the model with $\delta=0$ is the algebraic model~(\ref{pol}) 
for which the variation does not depend on the spacing $d$ (see dotted horizontal line in Fig.~\ref{figgcfmmd}, bottom right panel). 
Increasing $b$ emphasizes the impact of $K$. 
Yet, the proportion tends to the constant minimal value obtained when $b=0$ as $d$ increases for any $b>0$.  
Oppositely to the exponential model (see Fig.\ref{figuq}, left panel), the proportions do not tend to zero and 
the role of $K$ remains not negligible even if $d\rightarrow\infty$.

\section{Summary}\label{dis}

The general stability condition provided in Eq.~(\ref{CK}) is applied to a large class of car-following and pedestrian models 
based on relaxation processes. 
The stability systematically occurs when the relaxation times are sufficiently small. 
However, the precise nature of the influence of the number of predecessors in interaction $K$ on stability strongly depends on the type of models. 

The critical relaxation times converge as $K$ increases for the distance-based models with additive repulsion 
(as, e.g., the social force model \cite{Helbing1995}), 
for any repulsion function as soon as it tends to zero as the spacing increases. 
Therefore, the parameter $K$ has only limited effects on the stability.  
For this model class, a cut-off radius in the simulation has no influence on the dynamics if it is sufficiently large. 
The convergence describes a damped oscillation and the impact of the predecessor number on the stability 
is not systematically positive. 
This is not observed with optimal velocity models for which the stability always increases as the predecessor number increases 
(see, e.g., \cite{Lenz1999}). 
The role of the parameter $K$ on the stability is not negligible when the interactions are large   
(i.e.~distance-based models with repulsion $f(d)\propto e^{-dc}$ or $d^{-q}$ or OV model such that  
$a_k=\tau k^{-q}$ with small $c,q$). 
On the opposite, the stability condition weakly depends on $K$ for short-range interactions (i.e.~large $c,q$). 

The dynamics of the generalized centrifugal force $f(d,v,v_1)\propto \big(\gamma+\delta(v-v_1)\big)^2\big(d-b(v+v_1)\big)^{-1}$ is different. 
When the speed difference term is taken into account (i.e.~for $\delta=1$), the critical relaxation time $\tau_K$ diverges as $K$ increases. 
This means that the stability always holds for sufficiently high number of predecessors in interaction. 
Therefore the cut-off radius influences the dynamics and has to be calibrated as a parameter. 
The size slope parameter $b$ positively impacts the stability. 
However, $b$ tends to decrease the equilibrium speed.  
As for the distance-based models, the critical relaxation time $\tau_K$ converges as $K$ increases  
when the speed difference term is not taken into account (i.e.~for $\delta=0$). 
Yet the speed of convergence can be slow when $b$ and $d$ are high. 
Here the impact of $b$ on the stability can be either positive, when $K$ is high, or negative, when $K$ is low.

We summarize in Table~\ref{summ} where the models are distinguished according to:
1.~Their range, that can be short or long. 
2.~The definition of the mean speed in homogeneous configuration, that has in one dimension to be positive. 
3.~The influence of the number $K$ of predecessors in interaction. The influence can be bounded, unbounded, non-decreasing or 
non-monotonic.
4.~The impact of a cut-off radius on the dynamic. This is a consequence of the previous bounded influence of the predecessor number $K$. 
The cut-off is implicit when $K$ has limited effects (i.e.~bounded influence) while it has to be calibrated as a parameter for 
unbounded influence of $K$. 
5.~The most unstable wavelength. The most unstable wavelength can be the longest wave. 
In such a case the characteristic equation of the linear stability can be solved by expanding linear approximations 
(as in, e.g., \cite{Herman1959,Bexelius1968,Nagatani1999,Hu2014}). 
In some other cases the most unstable wavelengths have specific sizes and the characteristic equation has to be solved for any wavelength.

\begin{table}[!ht]\small\vspace{2mm}\begin{center}
 \begin{tabular}{c|c|c|c|c|c}
{Models} &Range&Positive equil. speed&Stability for $K$&Cut-off radius&Unstable wave \\[1mm]
\hline&&&&&\\[-1mm]
Exponential&Short&$A\tau\sum_{k=1}^Ke^{-kd/B}\le v_0$&Bounded,&Implicit&Longest\\[-.25mm]
SFM \cite{Helbing1995}&&&non-monotonic&&\\[1.5mm]
Algebraic&Long&$A\tau\sum_{k=1}^Kk^{-q}\le u^qv_0$&Bounded,&Implicit&Characteristic\\[-.25mm]
\cite{Guo2010,Helbing2000}&if $q\le 1$&&non-monotonic&&for $K$ large\\[1.5mm]
\hline&&&&&\\[-1mm]
OV \cite{Lenz1999}&Long&Given by $V(\cdot)$&Unbounded, &To be &Characteristic\\[-.25mm]
$q\le1$&&&non-decreasing&calibrated&for $K$ large\\[1mm]
$q>1$&Short&Given by $V(\cdot)$&Bounded, &Implicit&Characteristic\\[-.25mm]
&&&non-decreasing&&for $K$ large\\[1.5mm]
\hline&&&&&\\[-1mm]
GCF \cite{Chraibi2010}&Long&Solution of $K+1$&Unbounded,&To be &Longest\\[-.25mm]
$\delta=1$&&degree polynomials&non-decreasing&calibrated\\[1mm]
$\delta=0$&Long&Solution of $K+1$&Bounded,&Implicit&Characteristic\\[-.25mm]
&&degree polynomials&non-decreasing&&for $K$ large
 \end{tabular}\vspace{2mm}
 \caption{Summary of the influence of the number of neighbors in interaction $K$ on the stability for the exponential and algebraic, 
 the optimal velocity and the generalized centrifugal force acceleration-based models.}
 \label{summ}\end{center}
\end{table}

\section{Conclusion}\label{cl}

\begin{figure}[!ht]
\begin{center}
\includegraphics[width=.9\textwidth]{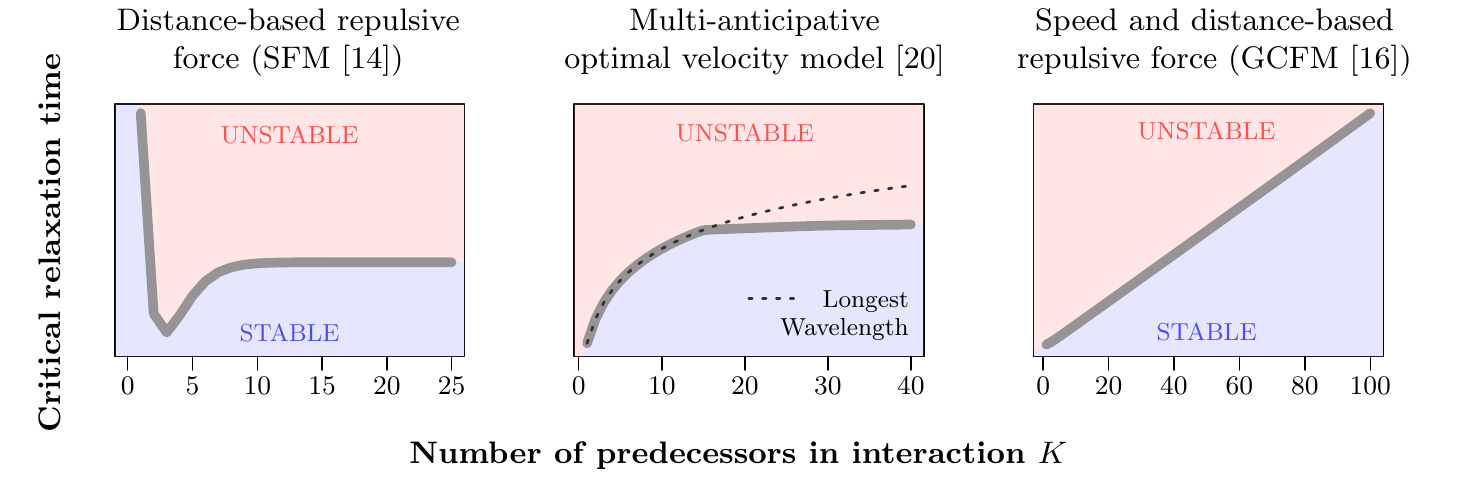}
\caption{Illustration of the stability condition for the three classes of models. 
Left panel: Exponential distance-based repulsive force. 
The stability improves not systematically as $K$ increases. 
Middle panel: OV model. 
The stability condition is bounded by unstable waves having a characteristic length. 
Right panel: GCF model. 
The stability linearly improves as $K$ increases.}
\label{figsumm}
\end{center}
\end{figure}

The linear stability conditions of uniform solutions are calculated in one dimension for any wavelength 
for a large class of acceleration-based microscopic flow models with $K\ge1$ predecessors in interaction. 
The framework is general and includes many pedestrian as well as road traffic models. 
The influence on stability of the number of predecessors in interaction 
depends strongly on the type of models. 
It can be both positive or negative, and also irrelevant in certain cases (see Fig.~\ref{figsumm}).

We observe that the increase of predecessors in the interaction not systematically gives increase of the stability. 
This result contrasts with the classical properties of OV traffic models 
for which increase of the interaction range systematically yields a stability improvement 
\cite{Bexelius1968,Lenz1999,Nagatani1999,Hoogendoorn2006,Zhu2007,Hu2014}, as well as to 
expected features of connected and autonomous vehicles \cite{Darbha1999,Wang2004b,monteil2014}. 

The stability condition converges as $K$ increases for a particular class of distance-based models that notably includes 
the social force model \cite{Helbing1995}.  
The parameter $K$ is only partially relevant in the dynamics and a cut-off radius during the simulation can be reasonably used. 
 
From a technical side, we point out that the most unstable wavelength is not systematically the longest wave. 
This is especially the case when an important number of predecessors is taken in account for the interaction. 
In such a case, linear approximations used 
to solve the characteristic equation of the linear stability \cite{Herman1959,Bexelius1968,Nagatani1999,Hu2014} are no more valid and 
the stability has to be analysed for any wavelength.

Note that the analysis is carried out in one-dimension and for strictly asymmetric interaction. 
Pedestrian flows and heterogeneous urban traffic stream in two dimensions, and with asymmetric interaction. 
Multi-dimensional streaming may have stability conditions different from the basic 1D streams and has to be investigated specifically. 
Furthermore, other speed-based models such that the full velocity difference model \cite{Jiang2001} remain to be analysed. 
The impact of the speed difference term on the stability may have different characteristics than 
the systematic stability improvement we observe with the GCF model. 
 
\section*{Acknowledgements}
The authors acknowledge financial support from the Deutsche Forschungsgemeinschaft (DFG) under grants SCHA 636/9-1 and SE 1789/4-1. 
\bigskip



\normalsize
\section*{Appendix 1: Matrix calculation of the linear stability condition}

The linear system (\ref{modlin}) is 
\be\dot{\mathbf Y}(t)=M\mathbf Y(t)\tag{A1}\ee 
with \be\mathbf Y(t)=T\left({\bar x}_1(t), {\bar v}_1(t),{\bar x}_2(t), {\bar v}_2(t),\ldots, {\bar x}_N(t), {\bar v}_N(t)\right),\tag{A2}\ee 
$T(\mathbf x)$ being the transpose of the vector $\mathbf x$, 
and, denoting $\varphi_K=-\sum_{k=1}^K \alpha_k$,
\be
M=\bmat
0&1&0&\ldots&&&&&&&&&\\
\varphi_K&\beta_0&\alpha_1&\beta_1&\ldots&\alpha_K&\beta_K&0&\ldots&&&&\\
0&0&0&1&0&\ldots&&&&&&&\\
0&0&\varphi_K&\beta_0&\alpha_1&\beta_1&\ldots&\alpha_K&\beta_K&0&\ldots&&\\[0mm]
&&&&&&&&&&&&\\[3mm]
&&&&&&&&&&\ldots&0&1\\
\alpha_1&\beta_1&\ldots&\alpha_K&\beta_K&0&\ldots&&&&&\varphi_K&\beta_0
\emat.
\tag{A3}\ee
The size of the vectors $\mathbf Y$ and $\dot{\mathbf Y}$ is $2N$, while $M$ is a $2N\times2N$ matrix. 
The solution of this linear system is 
\be\mathbf Y(t)=e^{Mt}\,\mathbf Y(0).\tag{A4}\ee
We have to check if $M$ is diagonalizable. 
If it is, we will have $M=P\mbox{diag}(\lambda_1,\ldots, \lambda_{2N})P^{-1}$ with $\mbox{diag}(\mathbf x)$ 
the diagonal matrix with coefficients $\mathbf x$. 
This allows to obtain 
\be e^{Mt}=P\,\mbox{diag}\big(e^{\lambda_1t},\ldots, e^{\lambda_{2N}t}\big)\,P^{-1}\tag{A5}\ee
and to calculate a solution of the system. 
The results show that the system converge to a stationary state if $M$ is diagonalizable with $\Re(\lambda_n)\le0$ for all $n$. 
We will determinate the conditions for that the system converges to the vector nil (i.e.~a uniform configuration is linearly stable).
The matrix $M$ has the form
\be
M=\bmat
M_0&M_1&\ldots&M_K&&\\
&M_0&M_1&\ldots&M_K&\\
&&&&&\\[4mm]
M_1&\ldots&M_K&&&M_0\\
\emat
\qquad\mbox{with}\qquad
\ba M_0&=&\bmat 0&1\\\varphi_K&\beta_0\emat,\\[5.5mm]
M_k&=&\bmat 0&0\\\alpha_k&\beta_k\emat,\quad k=1,\dots,K.
\ea
\tag{A6}\ee
The matrix is invariant per circular permutation of $2\times2$ block, 
i.e.~we have $\theta_2(M)=M$ by denoting the operator shifting all the coefficients of a matrix 
(or a vector) of 2 ranks to the right and 2 ranks to the down
\be
\begin{array}{ccccc}
\theta_2&:&M&\mapsto&\theta_2(M)\\
&&(m_{i,j})_{i,j}&\mapsto&(m_{i-2,j-2})_{i,j}.
\end{array}
\tag{A7}
\ee

If $\lambda\in\mathbb C$ is an eigenvalue associated to the eigenvector $\mathbf u\in\mathbb C^{2N}$, then 
\be M\mathbf u=\lambda\mathbf u.\tag{A8}\ee
We have  
\be\theta_2(M\mathbf u)=\theta_2(M)\,\theta_2(\mathbf u)=M\,\theta_2(\mathbf u)=\theta_2(\lambda\mathbf u)=\lambda\,\theta_2(\mathbf u).
\tag{A8}\ee
By construction, $\theta_2(\mathbf u)$ is also an eigenvector associated to $\lambda$. 
This implies 
\be\mathbf u=\gamma\theta_2(\mathbf u),\tag{A9}\ee 
with $\gamma\in\mathbb C$. 
Denoting $\mathbf u=T\left(u_1,u_2,\ldots, u_{2N}\right)$, we obtain 
\be T\left(u_1,u_2,u_3,\ldots, u_{2N}\right)=\gamma\times T\left(u_{2N-1},u_{2N},u_1,\ldots, u_{2N-2}\right)\tag{A10}\ee
and by iterating 
\be u_1=\gamma u_{2N-1}=\gamma^2 u_{2N-3}
=\ldots=\gamma^N u_1\quad\mbox{and}\quad u_2=\gamma u_{2N}=\gamma^2 u_{2N-2}
=\ldots=\gamma^N u_2.\tag{A11}\ee
This directly implies
\be
\gamma^N=1\quad\mbox{or}\quad \gamma=\sqrt[N]{1}\quad \mbox{i.e.} \quad \gamma=\gamma_l=e^{2i\pi l/N},\quad l=1,\ldots, N,
\tag{A12}\ee
as well as $u_{2m+1}=u_1\,\gamma_l^m$ and $u_{2m+2}=u_2\,\gamma_l^m$ for all $m=0,\ldots, N-1$. 
The eigenvectors are 
\be\mathbf u=u_1\,\mathbf u_l^{(1)}+u_2\,\mathbf u_l^{(2)},\tag{A13}\ee
 with $(u_1,u_2)\in\mathbb C^2$, 
$\mathbf u_l^{(1)}=T\left(1,0,\gamma_l,0,\gamma_l^2,0,\ldots, \gamma_l^{N-1},0\right)$ and $\mathbf u_l^{(2)}=T\left(0,1,0,\gamma_l,0,\gamma_l^2,\ldots, 0,\gamma_l^{N-1}\right)$. 
The eigenvalue $\lambda_l$ associated to the eigenvector $\mathbf u_l$ is the solution of $M\mathbf u_l=\lambda_l\mathbf u_l$ or again 
$\lambda_l u_1=u_2$ and $\lambda_l u_2=u_1\varphi_K+u_2\sum_{k=0}^K\beta_k\,\gamma_l^k+u_1\sum_{k=1}^K\alpha_k\,\gamma_l^k$. This is
\be
\lambda_l^2-\lambda_l\sum_{k=0}^K\beta_k\,\gamma_l^k+\sum_{k=1}^K\alpha_k(1-\,\gamma_l^k)=0.
\label{eqm}\tag{A14}
\ee
There are two distinct complex eigenvalues $\lambda_l^{(1)}$ and $\lambda_l^{(2)}$ associated to the eigenvector $\mathbf u_l$. 
Since $l$ varies from $1$ to $N$, there are $2\times N$ distinct eigenvalues for $M$. 
This proves by construction that $M$ is diagonalizable. 
The equation (\ref{eqm}) is the characteristic equation (\ref{eq}) with $\theta=2\pi l/N$.

For $l=N$, $\gamma_N=1$, $\lambda_N^{(1)}=\sum_{k=0}^K\beta_k$ and $\lambda_N^{(2)}=0$. 
If we assume that 
\be
\sum_{k=0}^K\beta_k<0\quad\mbox{and}\quad \Re\left(\lambda_l^{(p)}\right)<0\qquad
\left|\ba\mbox{for all}\quad p=1,2\\\mbox{and all}\quad l=1,\ldots, N-1,\ea\right.
\tag{A15}
\label{C}
\ee
then we have $\lim_te^{Dt}=\mbox{diag}(0,\ldots,0,1)$ and 
\be\lim_{t\rightarrow \infty}\mathbf Y(t)=\lim_{t\rightarrow \infty}Pe^{Dt}P^{-1}\mathbf Y(0)= 
T\Big(\sum_{n=1}^N \bar x_n(0),0,\ldots, \sum_{n=1}^N \bar x_n(0),0\Big),
\tag{A16}\ee
with $P=\big[ \mathbf u_1^{(1)}\ \mathbf u_1^{(2)}\ldots\ \mathbf u_N^{(1)}\ \mathbf u_N^{(2)}\big]$.
Therefore, under the assumptions (\ref{C}), $\mathbf Y(t)$ converges to the vector nil if
\be
\sum_{n=1}^N \bar x_n(0)=\sum_{n=1}^N x_n(0)-x^H_n(0)=0.
\tag{A17}\ee
This equality allows to identify the unique uniform configuration which is linearly stable for the system.

\section*{Appendix 2: Solved examples}
We illustrate here the linear stability condition (\ref{CK}) with two particular models for which the condition 
is well known in the literature. 

\paragraph{Model with one predecessor}

For $K=1$, the model is the second order  model with one predecessor in interaction  
\be
\left\{\ba
\dot x_n(t)&=&v_n(t),\\[1mm]
\dot v_n(t)&=&A\left(v_n(t),x_{n+1}(t)-x_n(t),v_{n+1}(t)\right)
\ea\right.
\label{mod2}\tag{A18}
\ee
investigated in \cite{Tordeux2012}. 
We suppose that for a mean spacing $d$, exists speed $v$ such that $A(v,d,v)=0$.
We have for this model the three parameters $a=\alpha_1$, $b=\beta_0$ and $c=\beta_1$, and, denoting $c_l=\cos(2\pi l/N)$ and $s_l=\sin(2\pi l/N)$, 
the parameters (\ref{pq}) are 
\be\left|\ba\mu_l=-\left(b+c c_l\right),\\[1mm]
\nu_l=a\left(1-c_l\right),\ea\right.\qquad
\left|\ba\sigma_l=-c s_l,\\[1mm]\rho_l=-a s_l.\ea\right.\tag{A19}\ee
The first condition in (\ref{CK}) is 
\be \mu_l=-\left(b+c c_l\right)>0.\tag{A20}\ee
This condition corresponds to the first one related in \cite[see Eq.~(B12)]{Tordeux2012}. 
It holds for all $l$ if $b+|c|<0$. 
This makes true the preliminary assumption (\ref{Cp}).
Assuming $a>0$, the second condition of (\ref{CK}) is
\be
\ba
\mu_l(\nu_l\mu_l+\rho_l\sigma_l)-\rho_l^2
=b^2-bc-a+c_l(bc-c^2-a)>0,\quad\forall l=1,\ldots, \lceil N/2\rceil.
\tag{A21}
\ea
\ee
This condition is the one related in \cite[Eq. (B19)]{Tordeux2012}. It holds for all $l$ if
\be
b^2-c^2-2a>0.
\tag{A22}
\ee

\paragraph{Optimal velocity model}

We investigate the multi-anticipative optimal velocity model~(\ref{mod3}) with $K\ge1$ predecessors in interaction \cite{Lenz1999}. 
With this model, the equilibrium speed $v$ corresponding to the mean spacing $d$ is $v=V(d)$, while $\beta_0=-\sum_{k=1}^Ka_k$, 
$\beta_k=0$, for all  $k\not=0$, and $\alpha_k=V'(d)/k$, for all $k$. 
Denoting $c_{lk}=\cos(2\pi lk/N)$ and $s_{lk}=\sin(2\pi lk/N)$ the parameters (\ref{pq}) are 
\be\left|\ba\mu_l=\sum_{k=1}^Ka_k,\\[1mm] 
\nu_l=V'(d)\sum_{k=1}^K\frac{a_k}k\left(1-c_{lk}\right),\ea\right.\qquad\left|\ba
\sigma_l=0,\\[1mm] \rho_l=-V'(d)\sum_{k=1}^K\frac{a_k}ks_{lk}.\ea\right.\tag{A23}\ee 
The first condition (\ref{CK}) is 
\be\sum_{k=1}^Ka_k>0.\tag{A24}\ee 
It is always true and implies the preliminary assumption (\ref{Cp}). 
The second condition (\ref{CK}) is, after rearranging
\be
0<V'(d)<\frac{\Big(\sum_{k=1}^Ka_k\Big)^2\sum_{k=1}^K\frac{a_k}k\left(1-c_{lk}\right)}{\left(\sum_{k=1}^K\frac{a_k}ks_{lk}\right)^2},\quad\forall l=1,\ldots, \lceil N/2\rceil.
\label{CKovm}\tag{A25}
\ee
This is the condition related in \cite[see Eq. (14)]{Lenz1999}. 
Note that the case $K=1$ corresponds to the well know optimal velocity model \cite{Bando1995}. 
For this model, the condition is 
\be V'(d)<a_1(1-c_l)/s_l^2=a_1/(1+c_l).\tag{A26}\ee 
Since $1/(1+c_l)>1/2$, the condition holds for all $l=1,\ldots, N-1$ if 
\be V'(d)<a_1/2\tag{A27}\ee (see \cite{Bando1995}).

\end{document}